\documentclass[twocolumn,groupedaddress,nofootinbib,aps,prl]{revtex4-1}
\usepackage{amsmath}
\usepackage{epsfig,color}
\usepackage{graphicx}
\usepackage{dcolumn}
\usepackage{bm}
\usepackage{multirow}
\usepackage{bm}
\usepackage{enumerate}
\usepackage{graphicx}
\usepackage{verbatim}
\usepackage{hyperref}
\usepackage[utf8]{inputenc}
\usepackage[english]{babel}
\usepackage{amssymb,color}
\usepackage{amsfonts}
\usepackage{graphicx}
\usepackage{dcolumn}
\usepackage{float}
\usepackage{bm}
\def\beq{\begin{equation}}
\def\eeq{\end{equation}}
\def\bea{\begin{eqnarray}}
\def\eea{\end{eqnarray}}

\begin{document}
\title{Phase transition in fluctuations of interacting spins at infinite temperature}

\author{V. N. Gorshkov}
\address{National Technical University of Ukraine, 37 Prospect Peremogy, Kyiv 03056, Ukraine}

\author{N. A. Sinitsyn}
\address{T-4, Los Alamos National Laboratory, Los Alamos NM 87545}

\author{D. Mozyrsky} \email{mozyrsky@lanl.gov}
\address{T-4, Los Alamos National Laboratory, Los Alamos NM 87545}

\date{\today}
\begin{abstract}
The high temperature limit of interacting spins is usually not associated with  ordering or critical phenomena. Nevertheless, spontaneous fluctuations of a local spin polarization at equilibrium have nontrivial dynamics even in this limit.  Here, we demonstrate that the spin noise power spectrum of these fluctuations can undergo discontinuous changes as a function of an external magnetic field.
As a simple illustration, we consider a model of Ising-like long range spin-spin interactions with a transverse magnetic field as a control parameter.
This system undergoes a phase transition associated with disappearance of the noise power peak  responsible for the most detrimental decoherence effect of the interactions.
\end{abstract}
\maketitle

A phase transition  is a discontinuous change of some measurable characteristic of a many-body system. Traditionally, phase transitions in condensed matter have been associated with sharp changes of a long range order at equilibrium when a control parameter changes across a critical value. The phase transitions are observed as divergence of linear response characteristics, such as the magnetic susceptibility near the critical point.

Any long-range ordering, however, is possible only at a sufficiently low temperature, $T$. Moreover, in the large temperature limit, the measurable linear response characteristics are vanishing because  the changes of microstate probabilities in the  equilibrium Gibbs distribution are suppressed  by the factor $1/(k_BT)$.

In this letter, contrarily to such expectations, we argue that interacting spin systems may show the basic features of a phase transition even in the large temperature limit at thermodynamic equilibrium. This critical phenomenon should be accessible for experimental studies by spin noise spectroscopy (SNS) \cite{SNS}, which probes dynamics of mesoscopic spin fluctuations at equilibrium. The basic measurable characteristic for the SNS is the noise power spectrum. It is formally related to the dynamic magnetic susceptibility, as articulated by the Fluctuation Dissipation Theorem \cite{SNS}. However, unlike the susceptibility, the power spectrum is not suppressed and has nontrivial features even at the infinite temperature.

For demonstration, we suggest Metal–Organic Frameworks  -- the organic materials that incorporate metallic ions with uncompensated spins in a regular array \cite{MOF}. The atomic spins can be placed at a sufficient distance from each other to remove exchange interactions. Thus, such spins form a macroscopic array of qubits with considerable coherence time. At cryogenic temperatures, $T<4K$,  the phonon-related mechanisms of decoherence in molecular spin qubits are  suppressed. The main sources of the decoherence are then hyperfine, of order of $\sim10-10^2$G, and magnetic dipole fields from the neighboring spin qubits ($\sim10^2-10^3$G)  \cite{mol-nano}. The hyperfine effects can be suppressed further by isotopic purification \cite{purified}, but the random long-range dipole fields represent an unsolved problem for quantum control. Thus, large qubit systems induce collective decoherence effects that are absent on the  level of a single qubit.

Here, we explore the coherent behavior of long-range interacting spins in a moderate external magnetic field. The most unfortunate effect of the dipole spin-spin interactions is the relatively slow collective  relaxation of the local spontaneous spin field fluctuations. The latter mis-align the net effective fields that act on individual spins from the external field axis. Therefore, there is no possibility to define a single rotation axis for the spins.

The main finding of our theory is that this most detrimental decoherence effect disappears at relatively week, of an order of the typical dipole field fluctuation, value of the external field. This is manifested as a discontinuous disappearance of one of the peaks in the noise power spectrum at a critical value of the external field. Our theory demonstrates other similarities with the conventional phase transitions, such as  divergence of the spin-lattice relaxation rate near a critical point.

As a theoretical model we consider the Hamiltonian with long-range Ising-like spin-spin interactions in a controllable transverse magnetic field, $f$:
\begin{equation}\label{1}
H = \sum_{i< j}  J_{ij} \sigma^z_i\sigma^z_j + f \sum_i \sigma^x_i  \,,
\end{equation}
where $\sigma_z^i$ and $\sigma_x^i$ are the $z$ and $x$ components for the $i$'th spin, $J_{ij}$ are coupling constants \cite{ab}.

The temperatures $T\sim1$K are  much larger, in energy scale, than the dipole fields. Hence, there is no spin ordering at the thermodynamic equilibrium, and the equilibrium spin density matrix is proportional to a unit matrix. The local spontaneous spin fluctuations create an uncompensated local spin polarization, whose dynamics is characterized by an ensemble averaged spin auto-correlation function,
\begin{equation}\label{5}
C(t)\equiv \frac{1}{2}\langle  \{\sigma^z_i(0),\sigma^z_i(t)\}\rangle,
\end{equation}
where $\{\ldots \}$ stands for anticommutator and the averaging is over the measurement results of the spin and over all possible initial spin configurations in the system at $t=0$. This correlator is
directly related to the noise power spectrum \cite{SNS-exp,SNS-exp-rev,SNS}:
\begin{equation}
\label{noiseP1}
    g(\omega)=\int dt C(t)e^{i\omega t}, \quad \omega>0.
\end{equation}

We will assume that $J_{ij}$ is long range and so each spin in the sum in Eq.~(\ref{1}) interacts with a large number of spins in its neighborhood. Therefore the effective magnetic field, acting on the $i$-th spin along z-direction,
\begin{equation}\label{3}
h_i =  \sum_{j\neq i} J_{ij}\sigma^z_j  \,,
\end{equation}
can be viewed as a slow (semi)classical variable.  This approximation was justified for the large temperature limit in \cite{QD1,QD2}. The  spins then follow the Larmor equation of motion:
\begin{equation}\label{2}
{\dot\sigma}_i =  {\bf\Omega}_i\times {\bf \sigma}_i\,,
\end{equation}
where ${\bf\Omega}_i$ is a 3-component vector, ${\bf\Omega}_i=(2f, 0, 2h_i)$; $h_i$  can be time-dependent and should be determined self-consistently.

If fields $h_i$ were constant (in time), the spin correlator would be the sum of a constant and an oscillating component:
\begin{equation}\label{7}
C^0(t) = \frac{h^2_i}{h^2_i+f^2}+\frac{f^2}{h^2_i+f^2}\cos \Big(2t\sqrt{h^2_i+f^2}\Big)  \,.
\end{equation}
The Fourier transform~(\ref{noiseP1}) would have then two  delta-peaks: at $\omega=0$ and at $\omega= 2\sqrt{h^2_i+f^2}$. The spectral weight $h^2_i/(h^2_i+f^2)$ of the $\omega=0$ peak represents the fraction of spin polarization that does not experience the relaxation, whereas $f^2/(h_i^2+f^2)$ is  associated with the Larmor precession around ${\bm \Omega}_i$.

Since $h_i$ are different for different $i$, one should average Eq.~(\ref{7}) over $h_i$. The width of the Larmor peak would then correspond to the standard inhomogeneous broadening. The averaging over $h$-fields, however, would only modify the overall spectral weight of the zero-frequency peak, without any broadening to it.

In Fig.~1 we show the results of our numerical simulations for a spin chain with long-range interactions  in  Eq.~(\ref{7}) \cite{suppl}.
Figure~1 confirms the presence of  both noise power peaks for relatively small $f$. However, for various choices of the coupling distribution, inevitably,  at the external field above some critical value, the zero-frequency peak  disappears completely.
\begin{figure}[t!]
\includegraphics[width=1.1\columnwidth,keepaspectratio]{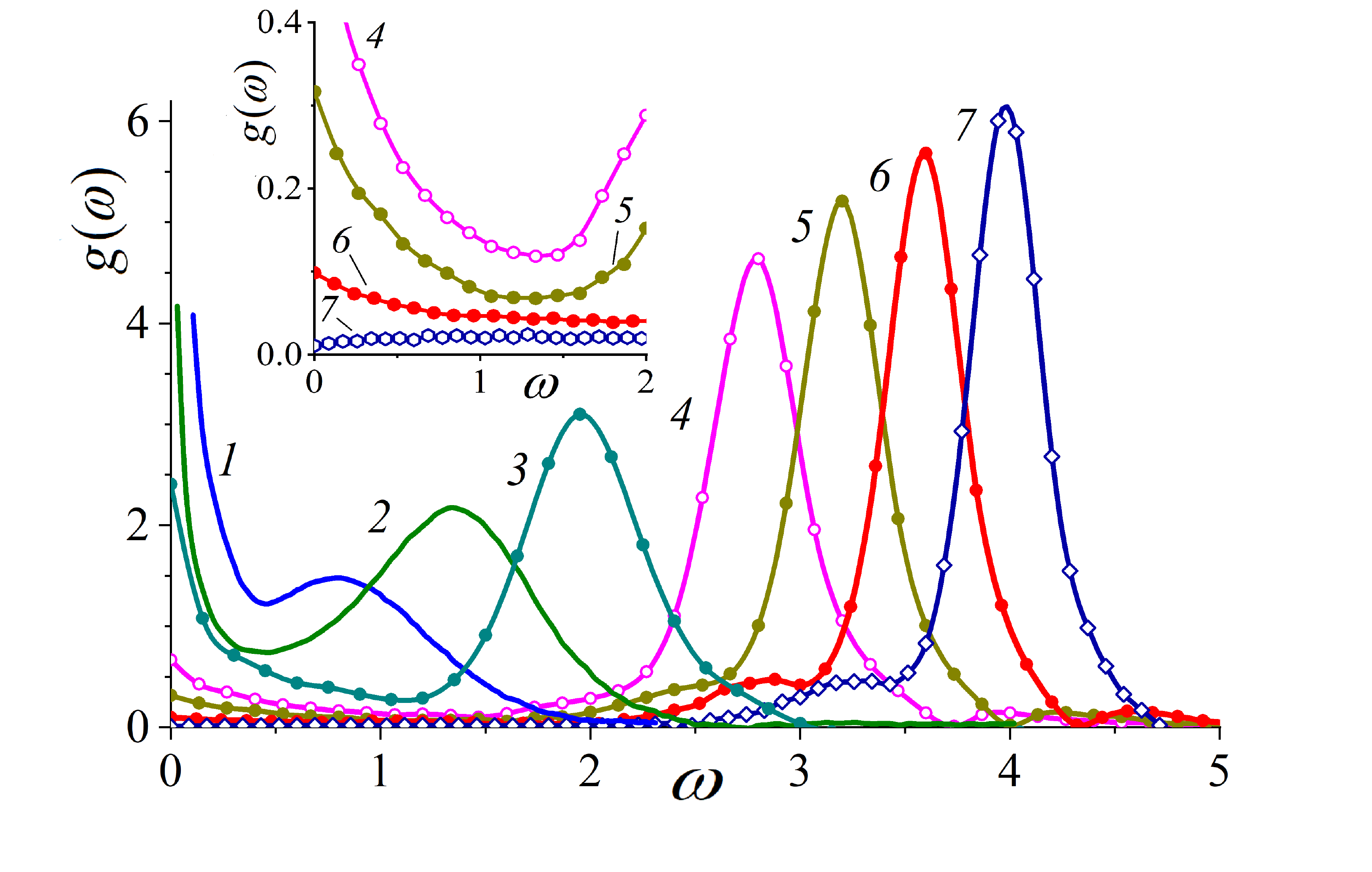}			
\caption{\label{fig1} The numerically generated (see supplementary file \cite{suppl} for details) spectrum of the spin auto-correlation function in the spin chain at different transverse fields $f$.  Curve 1: $f=0.5$; Curve 2: $f=0.75$; Curve 3: $f=1.0$; Curve 4: $f=1.4$; Curve 5: $f=1.6$; Curve 6: $f=1.8$ ; Curve 7: $f=2.0$. For all curves $\sum_j J_{ij}^2 = W^2/4$, where $W=3.62$.} 
\end{figure}

This observation was not expected because within the simple picture in which the fields $h_i$ are static, the zero-frequency peak does not disappear even in very large external fields. For $f\rightarrow \infty$, the ratio of the weights of the delta-peaks from (\ref{7}) decays as a power law $\sim h_i^2/f^2$. Hence, it seems that in order to just suppress the correlated lattice effects, the external field $f$ must be considerably larger than the equilibrium root-mean-square of $h_i$ at $f=0$. However, Fig.~1 shows that the zero-frequency peak disappears at some critical $f=f_c$, which is comparable to the typical $|h_i|$.

This  resembles the behavior  of a magnetization near the critical point of a ferromagnetic phase transition. The role of the order parameter in our case is played by the weight of the zero-frequency peak, which is the noise power integrated over the frequency range where this peak has noticeably nonzero values. This weight is finite for $f<f_c$ but is zero otherwise.

The analogy with conventional phase transitions can be enhanced by considering behavior of this peak's spectral weight for $f$ near $f_c$. The field $h_i$ in Eq.~(\ref{7}) has a quasi-static and a quickly fluctuating, due to fast spin rotations, components.
The quasi-static one, $h$, varies from one spin to another randomly. If  $J_{ij}$ are long range, then the distribution function $P(h)$ for the $h$-fields can be approximated by a Gaussian,
\begin{equation}\label{8}
P(h) = \frac{1}{\sqrt{2\pi h_0^2}}\ e^{-\frac{h^2}{2h_0^2}}  \,,
\end{equation}
where $h_0^2$ is the variance of the quasi-static interaction field. On the other hand, from the definition of $C(t)$ and  $h_i$, Eqs.~(\ref{5}, \ref{3}), and the interpretation of the first term in~(\ref{7}) as the quasi-static part of the field contribution to the noise power, we find that $\langle h_0^2 \rangle$ is expressible via the weight of the zero-frequency peak:
\begin{equation}\nonumber
h_0^2=\sum_{j,\,j\neq i} J_{ij}^2\, \frac{h^2_j}{h^2_j+f^2} \,,
\end{equation}
where we  used the fact that there are no same-time correlations between different spins at high temperatures. This leads to a self-consistency equation on $h_0$:
\begin{equation}\label{9}
\frac{W^2}{4}\, \int dh P(h)\, \frac{h^2}{h^2+f^2} = h_0^2,
\end{equation}
where  $W^2\equiv 4\sum_{j\neq i} J_{ij}^2$.
\begin{figure}[t!]
\includegraphics[width=1.0\columnwidth,keepaspectratio]{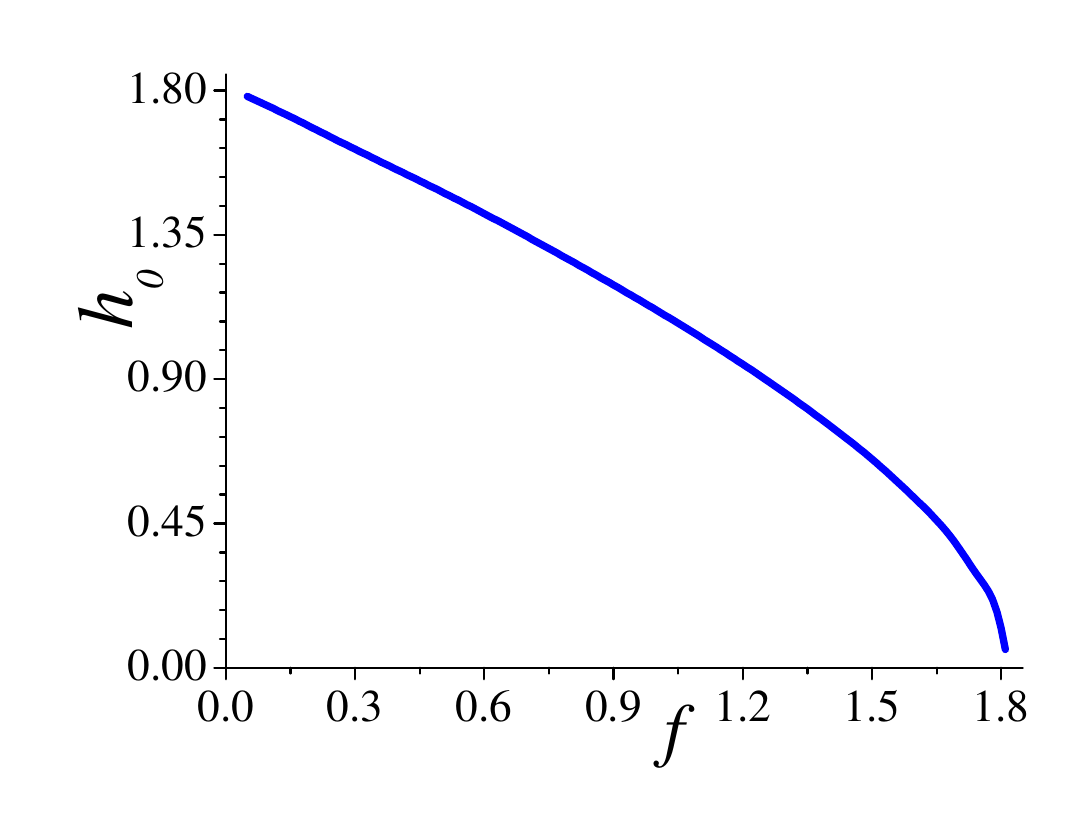}		\caption{\label{fig1} Numerical solution of  Eq.~(\ref{9}) for $W=3.62$. }
\end{figure}
Equation~(\ref{9}) can be readily solved numerically. It has always a trivial solution $h_0=0$, but for sufficiently small $f$ it also has a non-zero solution which is shown in Fig.~2. For $f\rightarrow 0$, $h_0\rightarrow W/2$, which is the typical spin-spin field fluctuation at $f=0$. As $f$ increases, the distribution $P(h)$ narrows until its width $h_0$ is $0$. This point corresponds to the critical field: $ f_c=W/2$.
For values of $f>f_c$, only the trivial solution exists, whereas at slightly smaller $f$ (than $f_c$), the non-zero solution is
\begin{equation}\label{delta}
h_0=f_c\sqrt{\delta/3}, \quad \delta= (f_c^2-f^2)/f_c^2 \,.
\end{equation}
This (square root) behavior is consistent with the presence of a 2nd order phase transition, with $h_0$ being the order parameter.
The self-consistency equation yields $f_c\simeq 1.81$. This result is in a very good agreement with numerical simulations: Comparing such estimates with Fig.~1,  the transition takes place between curves 6 and 7, which correspond to the values of $f=1.8$ and $f=2.0$, respectively.

The disappearance of the zero-frequency peak does not mean that the spin interaction effects are not present. For $f>f_c$, they are still manifested in the broadening of the Larmor peak. However, this effect is very different from the mis-aligning of quasi-static fields. The spin interaction effects resemble  now the effect of a fast noise that produces Lorentzian-like broadening of the Lamor precessions around the external field direction.

{\it Broadening of the zero-frequency peak:}  The delta-peak at $\omega=0$ associated with the first term in Eq.~(\ref{7}) has zero-width, which is not the case with the curves in Fig.~1. This broadening is the consequence of the dynamics of $h_i$. Thus, the phase transition and the broadening of the delta-peak have the same origin.

Not only the zero-frequency peak  looses power with growing $f$ but also it flattens as $f\rightarrow f_c$ (Fig.~1). This suggests that the width of this peak has critical behavior near the phase transition as well.
To confirm this, we rewrite Eq.~(\ref{2}) as an equation that contains only $\sigma^z_i$ components of the spins. For that we solve the equations containing the time derivatives of $\sigma^x_i$ and $\sigma^y_i$ in terms of $h_i$ and $\sigma^z_i$ and substitute these solutions (for $\sigma^x_i$ and $\sigma^y_i$) into the remaining equation for $\sigma^z_i$. The result is
\begin{equation}\label{4}
{\dot\sigma^z_i} = -4f^2 \int_0^t dt^\prime \sigma^z_i(t^\prime)  \, \cos\Big[2\int_{t^\prime}^t d\tau h_i(\tau)\Big]  \,,
\end{equation}
where $h_i$ is given by Eq.~(\ref{3}).

We separate the field $h_i(\tau)$ in Eq.~(\ref{4}) into low frequency (slow) and high frequency (fast) components:
\begin{equation}\label{10}
h(\tau) = h_s + h_f(\tau) \,,
\end{equation}
with $h_s$ being nearly a constant (yet, a random number distributed according to the distribution $P(h_s)$, e.g. Eq.~(\ref{8})) and $h_f(\omega)$ having non-zero Fourier components only at $|\omega|\ge 0$. That is, the power spectrum of the latter, $g_f(\omega)=\langle|h_f(\omega)|^2\rangle$, can be associated with the part of the spectrum (e.g. Fig.~1) that includes the peak at finite (i.e. Larmor) frequencies. This part corresponds, approximately, to the Fourier transform of the second term in Eq.~(\ref{5}), averaged over the distribution $P(h_s)$ \cite{suppl}.

To obtain the low frequency spin  dynamics, we apply the same frequency separation to $\sigma^z_i$, as we did for $h$ in (\ref{10}). Then, when writing Eq.~(\ref{4}) for a slow spin (in the following denoted by $\sigma_s^z$), we can average over the rapidly varying field $h_f$. By virtue of the central limit theorem the latter can be assumed to be a Gaussian variable (in the limit of long range coupling), and therefore, upon averaging over $h_f$ and assuming that $h_s$ is a constant, the cosine in Eq.~(\ref{4}) transforms into a function of $t-t^\prime$; (see  \cite{suppl} for explicit details). The resulting equation can be readily solved by the Laplace transform and one obtains the fluctuation spectrum of slow spins, $g_s(\omega, h_s)=\langle|\sigma_s^z(\omega,h_s)|^2\rangle$. In the supplemental materials we show that spectrum $g_s(\omega, h_s)$ is a Lorentzian, with the width being a function of the slow field $h_s$ \cite{suppl}.

The spectrum of the spin autocorrelation function at sufficiently small frequencies can be obtained by averaging $g_s(\omega, h_s)$ with respect to the distribution of slow fields $h_s$ according to Eq.~(\ref{8}), and accounting for that the slow spins constitute a fraction $h_s^2/(h_s^2+f^2)$ of all spins, e.g. the first term in Eq.~(\ref{7}),
\begin{equation}\label{14}
g(\omega) = \int dh_s P(h_s)\,\frac{h_s^2}{f^2+h_s^2}\, g_s(\omega, h_s) \,.
\end{equation}

{\it Noise spectrum at small $\omega$ near the transition point:} Since at the transition point (i.e., for $f\rightarrow f_c$) the order parameter $h_0$ vanishes (i.e., $P(h_s)$ approaches delta-function), we can evaluate the spectrum $g(\omega)$ by setting $h_s=0$ in $g_s(\omega, h_s)$. Evaluation of the Lorentzian $g_s(\omega, 0)$ is presented in the supplemental materials \cite{suppl}, and we quote the final result
\begin{equation}\label{17}
g(\omega) = \frac{\delta}{3}\,\frac{2\sqrt{2\pi}f_c}{\delta^2\omega^2+2\pi f_c^2} \,.
\end{equation}

Equation~(\ref{17}) clearly has properties consistent with those for the numerical power spectra presented in Fig.~1: as $f$ approaches $f_c$, the integrated weight of the spectrum in the  interval between $0$ and a frequency of the order of $O(f)$ is of the order of $\delta\ll 1$. Furthermore, as $f$ approaches  $f_c$, the spectrum of the slow spins flattens and, at the critical point, becomes equal to $0$ at the final interval at sufficiently low frequencies. Barring a contradiction of terms, the relaxation rate, $\sqrt{2\pi}f_c/\delta$,  diverges for slow spins. This is formally opposite to the divergent correlation length in case of conventional 2nd order phase transitions.

{\it Noise spectrum at small $\omega$ for $f\ll f_c$:} At sufficiently small $f$, the typical values of $h_s$ greatly exceed $f$ and therefore the relative fraction of slow spins, i e. the fraction in the integrand in Eq. (\ref{14}) can be replaced by $1$. The width $\Gamma$ of the Lorentzian $g_s(\omega, h_s)$ is evaluated in the supplemental materials \cite{suppl}: It turns out to be a strong (Gaussian) function of $h_s$. Then,  changing integration variable in Eq. (\ref{14}) from $h_s$ to $\Gamma(h_s)$, with {\it logarithmic accuracy} we obtain
\begin{equation}\label{18}
g(\omega)\sim\int_0^{\Gamma(0)} d\Gamma \,\frac{\Gamma^\alpha}{\omega^2 + \Gamma^2}\,,
\end{equation}
where $\alpha=(W^2-4h_0^2)/4h_0^2$ and $\Gamma(0)\sim f^{3/2}/W^{1/2}$; see \cite{suppl} for the details of derivation.

We see that for $\alpha\le 1$ and $|\omega|\ll\Gamma(0)$, Eq. (\ref{18}) yields  $g(\omega)\sim |\omega|^{\alpha-1}$. This power law dependence of the fluctuation spectrum is consistent with the shapes of the spectra obtained by numerical simulations in Fig. 1. Indeed, for $f\ll W$, $\alpha\simeq \sqrt{2\pi} f/W$, and therefore the power low dependence (with the exponent $\alpha-1$) suggests that the spectra with smaller $f$ tend to $\infty$ as $|\omega|\rightarrow 0$ more steeply than those with greater $f$. This is exactly the case with the curves 1 and 2 in Fig. 1.

Apart from the phase transition, at $\alpha=1$ the system exhibits a {\it crossover}: For $\alpha < 1$, the fluctuation spectrum is divergent at zero frequency, which indicates that some spins/spin domains (with sufficiently high $h_s$) do not evolve and remain frozen indefinitely. For $\alpha > 1$, Eq.~(\ref{18}) suggests that such picture breaks down (so that all spins have finite relaxation time). This happens when $h_0=W/\sqrt{8}$, which corresponds to $f\simeq 0.9$ (for $W=3.62$), e.g. Fig.~2. This value is clearly in agreement with what we observe in Fig.~1: While at $\omega=0$ the spectrum for the curve 2 with $f=0.75$ is divergent, it is finite for curve 3 with $f=1$.

The origins for the phase transition and the crossover can be understood if we consider a model with infinite range coupling ($J_{ij}=J\sim 1/\sqrt{N}=const$) \cite{LMG}. This model is integrable and, although missing the effects of disorder, provides an opportunity for a detailed, nearly exact analysis of the critical behavior. In the supplemental materials \cite{suppl} we show that the dynamics of the effective field $h$ (i.e. magnetization) in this model corresponds to the dynamics of a particle moving in a Ginzburg-Landau-like potential. The phase transition can be related to the emergence/disappearance of two potential minima/wells, each corresponding to a fixed point for $h$.
On the other hand, the crossover to the state with frozen magnetization is related to the existence of a classical trajectory that connects between these two wells. That is, for sufficiently small values of the transverse field $f$, the ``particle'' is bound to travel within a single well, while for $f$ higher than a certain threshold value, it can transition between the two wells.

{\it In conclusion,}  direct measurements of the spin fluctuations can reveal unusual dynamic phase transitions that have no analogs in the conventional theory of critical phenomena.
Our phase transition is associated with the disappearance of the low frequency fluctuations at certain (critical) value of the applied magnetic field. Also, at somewhat lower values of the field the system exhibits a crossover, which can be associated with the emergence/disappearance of very slowly relaxing states (i.e. sufficiently large spin domains) related to the existence of the dynamically stable fixed points for system's magnetization. \cite{suppl}.

The study of similar phase transitions in many-body fluctuations should be of practical importance. In addition to the~qubit control applications, such critical points  reveal nonperturbative, including topological, changes in the energy functional even though the system does not follow its mimimum. It should be insightful to relate such phenomena  to the properties of the ordered phases that emerge in the same systems at finite temperatures, and thus  predict the phase diagrams using only easy-to-detect high-temperature noise characteristics.

\begin{acknowledgements}
{\it Acknowledgements:}
We thank A. Balatsky, R. Malla and I. Martin for valuable discussions. The work at LANL is supported by LDRD program.
\end{acknowledgements}


\clearpage

\begin{widetext}
\section{Supplementary material for ``Phase transition in fluctuations of interacting spins at infinite temperature"}
\bigskip
\end{widetext}




\section{Numerical simulations}

We have carried out  numerical simulations for the system of $1000$ classical spins on 1-dimensional ring (i.e., with periodic boundary conditions) by preparing spins with random initial $\sigma^z_i(0)=\pm 1$ at equal weights (e.g. high temperature limit), and with $\sigma^x_i(0)=\sigma^y_i(0)=0$. We found that the choice of $J_{ij}$ does not  change the behaviour of the correlation function qualitatively as long as the spin-spin coupling is sufficiently long range, e.g. Figs.~1-3 of this supplementary. We also calculated the auto-correlation functions for the effective fields $h_i(t)$, which show behaviour very similar to the spin auto-correlation functions, signalling that  correlations between different spins are relatively weak, (see, e.g., the insets in Figs.~1-3). Furthermore, we observe in Figs. 1-3 that at small frequencies the behaviour of the spin auto-correlation functions for different choices of spin-spin couplings (i.e. red, blue and green curves) are very close even at the quantitative level. This is in agreement with out theoretical expectations that the small frequency behaviour of the fluctuation spectra is  controlled by the quantity $\sum_j J_{ij}^2$, rather than by the specific functional choice of couplings between spins.

In Fig.~1 of the main text, we present the Fourier transform of the correlation function, $g(\omega)$, at different values of $f$. For relatively small  $f$, e.g. curves 1-5, the correlation function exhibits two peaks: one at zero frequency, and another at, roughly, at Larmor frequency $2f$. While the nature of the latter peak is self-evident, the origin and the relevant physics of the former, zero-frequency peak is less obvious. We emphasize that the aforementioned behaviour occur for rather arbitrary long range couplings.

We argue (see also the main text) that the peak at $\omega=0$ is related to the existence of random domains of spins initially having the same z-component (i.e., $\pm 1$) and thus having relatively low relaxation times. Indeed, spins in such domains would ``see'' sufficiently strong effective fields $h_i$ and, therefore, would be less susceptible to transitions due to the transverse field $f$. That is, when $h_i\gg f$, the individual spins in such  domains would tend to align along $z$ direction rather than rotate along total magnetic field (of magnitude $2\sqrt{h_i^2+f^2}$; see more discussion below). In Fig. 4  we present ``screenshots'' for the time evolution of spin's z-components showing the existence of such ``metastable'' domains that confirm this argument.

\begin{figure}[t!]
\includegraphics[width=1.0\columnwidth,keepaspectratio]{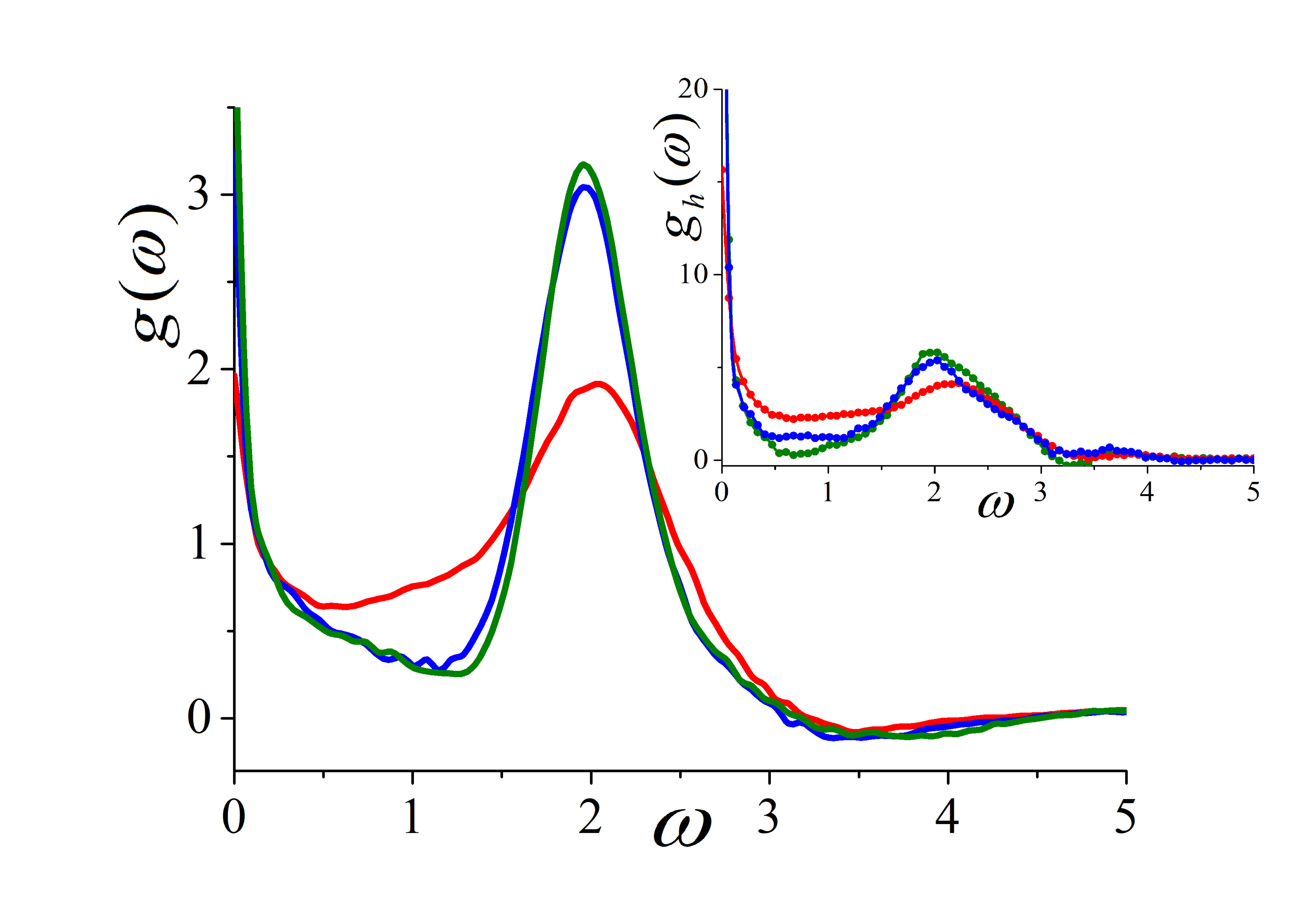}			
\caption{\label{fig1} Numerical simulations: Spectrum of spin auto-correlation function corresponding at $f=1.0$ for different couplings; Inset: Spectrum of the effective field auto-correlation function $C_h(t)=(1/2)\langle h_i(0)h_i(t) + h_i(t)h_i(0)\rangle$, with $h_i=\sum_j J_{ij}\sigma^z_i$. Here $g_h(\omega)=\int d\omega e^{i\omega t}\,C_h(t) $. For red curve $J_{ij}=1/|i-j|$, for blue curve $J_{ij}= a/|i-j|^2$ and for the green curve $J_{ij}= \exp [(1-|i-j|)/b]$. The values of coefficients $a$ and $b$ are chosen to satisfy condition $\sum_j J_{ij}^2 = W^2/4$, where $W=3.62$ for all three curves.}
\end{figure}
\begin{figure}[t!]
\includegraphics[width=1.0\columnwidth,keepaspectratio]{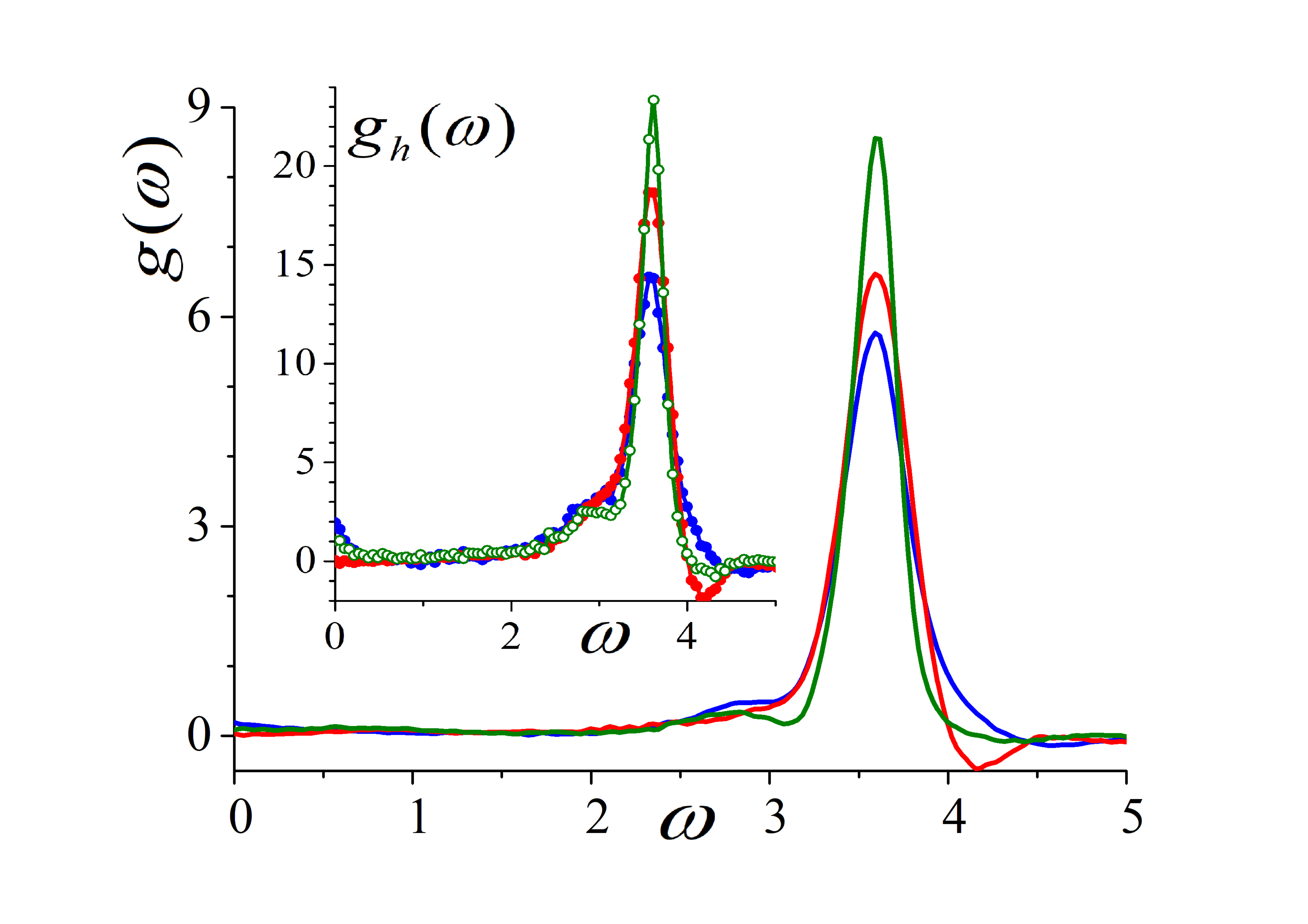}			
\caption{\label{fig2} Numerical simulations: Same as in Fig.~1, but for $f=1.8$.}
\end{figure}
\begin{figure}[t!]
\includegraphics[width=1.0\columnwidth,keepaspectratio]{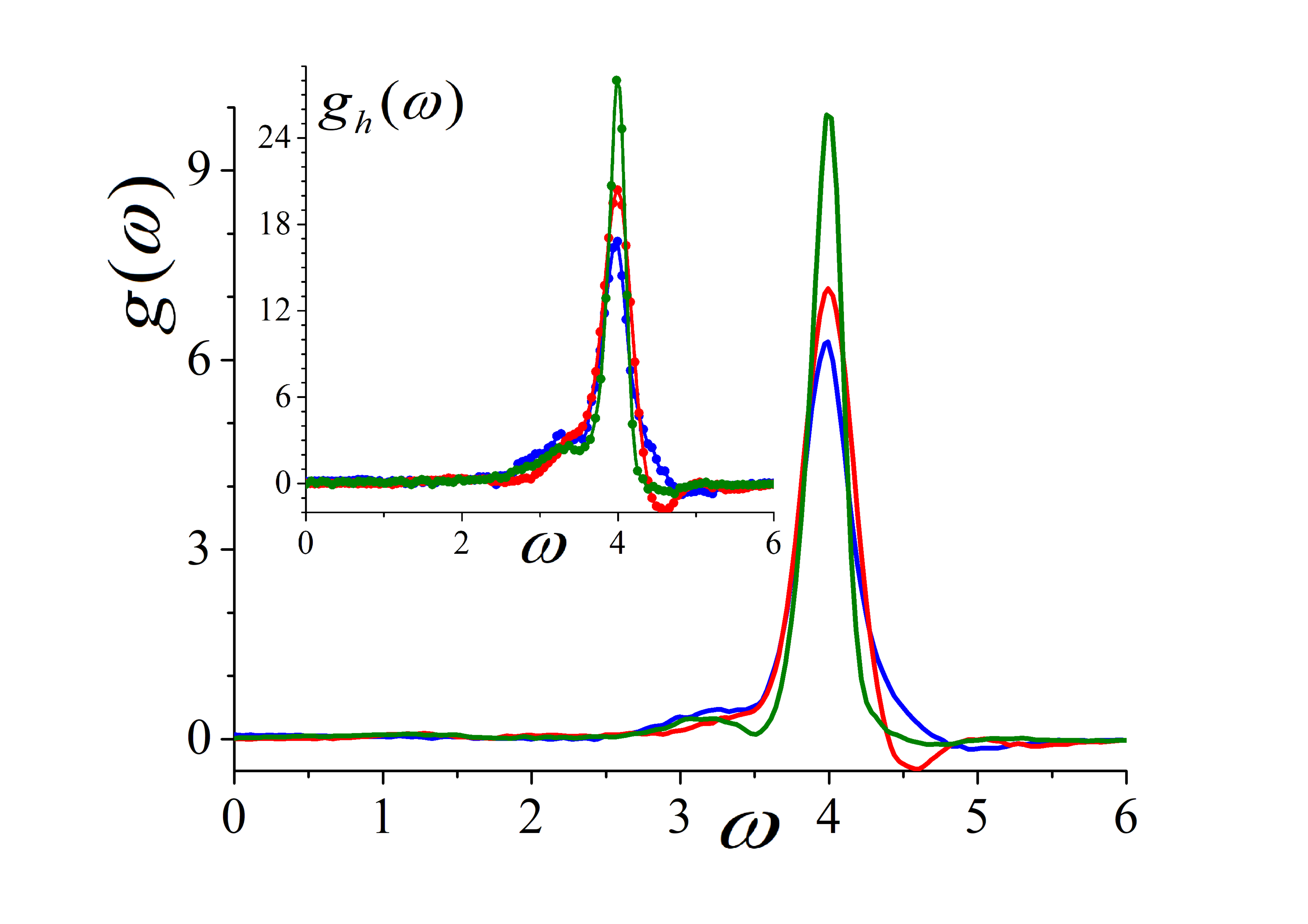}			
\caption{\label{fig3} Numerical simulations: Same as Figs.~1~and~2, but for $f=2.0$.}
\end{figure}
\begin{figure}[t!]
\includegraphics[width=1.0\columnwidth,keepaspectratio]{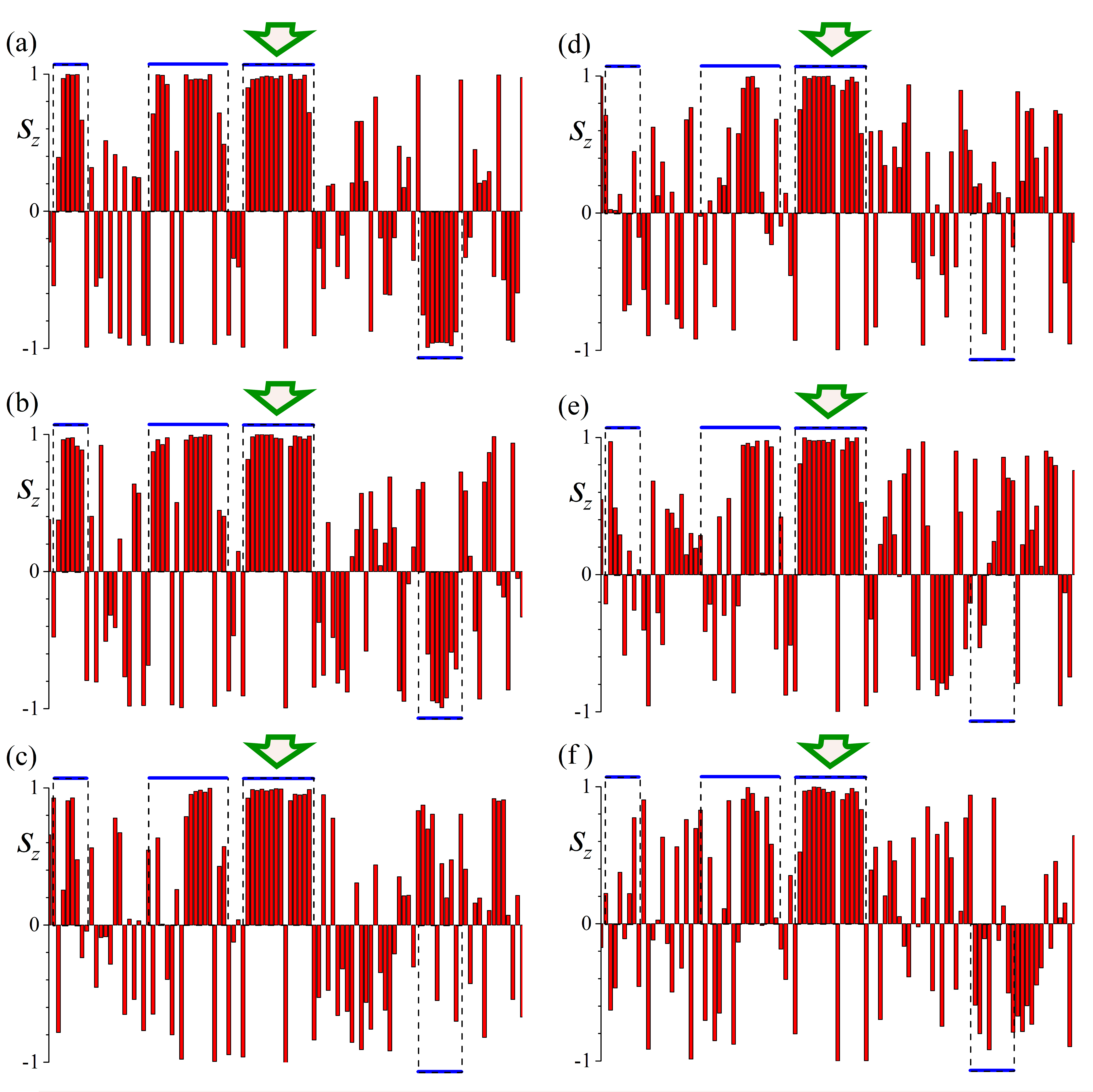}			
\caption{\label{fig4} Snapshots for simulated time-evolution of local spin $z$-polarization in a 1D spin chain with long-range interactions for $f=0.5$ and $J_{ij}=1/|i-j|$. The horizontal axis is the spin index.
The relaxation time is strongly dependent on the spin-fluctuation size. The relaxation time for large  domains is anomalously long and exceeds the entire accessible simulation time. Green arrow points to such a stable spin domain. }
\end{figure}

At higher transverse fields $f$, between curves 6 and 7 (for $f=1.8$ and $f=2.0$, respectively), the peak at $\omega = 0$ disappears, which indicates that a certain transition is taking place.  This transition is related to the existence of local, nearly constant in time, but random in space effective fields (along z-direction) that persists below some critical value of the transverse field $f$, and vanish above this critical value. Therefore we believe that the system exhibits a dynamical phase transition between the two regimes, with the root-mean-squared average of such fields being an order parameter for such transition.


\section{Interpretations of the dynamic phase transition: Lipkin-Meshkov-Glick model}

The Lipkin-Meshkov-Glick (LMG) model \cite{LMG} is a simplified version of our model, which can be understood in great detail. It is obtained from our model after an assumption of all-to-all equal interactions of $N$ spins:

\begin{equation}
H=(J/2) \left( \sum_{k=1}^N \sigma_k^z \right)^2+ f\sum_{k=1}^N \sigma_k^x
    \label{equal1}
\end{equation}
The  spin projections
\begin{equation}
S_{x,z}\equiv\frac{1}{2} \sum_{k=1}^N \sigma^{x,z}_k
\label{s}
\end{equation}
commute with the net spin operator $S^2$. Therefore, The Hamiltonian (\ref{equal1}) splits into invariant sectors with all possible values of $S=0,\ldots,N/2$, which have different multiplicity. The Hamiltonian for the sector with spin $S$, up to a constant, is given by the LMG model:
\begin{equation}
  H_{LMG} = 2JS_z^2 + 2fS_x.
    \label{LMG1}
\end{equation}
In the large temperature limit, all sectors with different $S$ will contribute to the noise power spectrum. However, for $N\gg 1$,
the main contribution will be produced by the sectors with typical spin fluctuation size
\begin{equation}
\sqrt{\langle S_z^2\rangle}  = \sqrt{N}/2 \gg 1\,,
\label{typical}
\end{equation}
where we have used Eq. (\ref{s}) and the fact that different individual spins are uncorrelated at the same time.

Let us introduce spherical coordinates $(\theta,\phi)$ that parameterize the large-spin state, so that
\begin{equation}
S_z=S\cos \theta, \quad S_x
=S\sin \theta \cos \phi.
\label{comps}
\end{equation}
Comparing Eq.~(\ref{comps}) with Eq.~(\ref{typical}), we notice that the value of $S$ can be related with the total number of spins $N$ through the relation
\begin{equation}
S^2\times \frac{1}{2}\simeq \frac{N}{4}\,,
\label{sn}
\end{equation}
so that the typical values of $S$ are of the order of $\sqrt{N/2}$. That is, in the limit of large $N$, the calculation of the correlation function for the z-component of spin magnetization, $\langle S_z(0)S_z(t)\rangle$, formally reduces to distribution of $S$'s according to Gaussian distribution $P(S)$ with variance $\sqrt{N/2}$, finding trajectories $S_z(t)$ for a fixed value of $S$ (corresponding to the initial conditions $S_x(0)=S_y(0)=0$ and $S_z(0)=S$), and then averaging the quantity $S_z(t,S)$ with respect to $P(S)$.

Eq.~(\ref{sn}) requires a comment. One could have argued that the coefficient $1/2$ in the left hand side of Eq.~(\ref{sn}) should instead be $1/3$, which would correspond to the angles $\theta$ and $\phi$ being uniformly distributed. Indeed, under such assumption the average of $S_z^2$ is equal to $1/3$:
$$
\langle S_z^2\rangle = \frac{1}{4\pi} \int d\,\Omega S^2\cos^2\theta = \frac{S^2}{3}\,,
$$
where $d\Omega= d\phi d\cos\theta$.
This, however, is not the case in the present calculation: Here we are interested in the correlation function of the spin magnetization, and, therefore, the initial magnetization is constrained by the  condition $S_x=S_y=0$, with $S_z$ being random. Then, the appropriate averaging corresponds to treating angle $\theta$ as a random number, uniformly distributed in the interval between $0$ and $\pi$. This gives factor $1/2$ in the left hand side of Eq.~(\ref{sn}).

The classical spin dynamics is found by minimizing the action with an effective Lagrangian \cite{LL1}
$$
L=S\dot{\phi}\cos \theta -H_{LMG}(\theta,\phi),
$$
where in our case
$$
H_{LMG}(\theta,\phi) = 2JS^2 \cos^2 \theta +2Sf\sin \theta \cos \phi.
$$

The equations of motion read:
\begin{equation}
{\dot \theta} = -2f \sin\phi\,,
\label{eqmo1}
\end{equation}
\begin{equation}
\sin\theta{\dot \phi} = 4JS\cos\theta\sin\theta -2f \cos\theta\cos\phi\,.
\label{eqmo2}
\end{equation}
After straightforward manipulation Eqs.~(\ref{eqmo1},s\ref{eqmo2}) can be cast in the following form:
\begin{equation}
\frac{d}{dt}(\sin\theta\cos\phi) = -4JS\cos\theta\sin\theta \sin\phi\,,
\label{eqmo3}
\end{equation}
\begin{equation}
\frac{d}{dt}(\sin\theta\sin\phi) = 4JS\cos\theta\sin\theta \cos\phi-2f\cos\theta\,.
\label{eqmo4}
\end{equation}

Using Eq.~(\ref{eqmo1}),  we rewrite Eq.~(\ref{eqmo3}) as
\begin{equation}
\frac{d}{dt}\big(f\sin\theta\cos\phi + JS\cos^2\theta\big) = 0\,,
\label{cons}
\end{equation}
which is nothing but the energy conservation, i.e., $H_{LMG}(\theta,\phi)=const$ along a trajectory. Using the initial conditions, $\cos\theta=1$ and $\sin\theta=0$ at $t=0$, Eq.~(\ref{cons}) can be integrated to give the relation between $\phi$ and $\theta$:
\begin{equation}
f\cos\phi = JS \sin\theta\,.
\label{rel}
\end{equation}

Eliminating $\sin\theta$ and $\cos\theta$ in Eq. (\ref{eqmo4}) and using  Eqs.~(\ref{eqmo1}, \ref{rel}), Eq.~(\ref{eqmo4}) can be rewritten as:
\begin{equation}
\frac{d^2\cos\theta}{dt^2} = 4\cos\theta\Big[2(JS)^2(1-\cos^2\theta)-f^2\Big]\,.
\label{eqmo5}
\end{equation}
Equation~(\ref{eqmo5}) has the desired form: it is a closed equation for the z-component of the spin magnetization $S_z=Scos\theta$.

Denoting $\cos\theta = z$, we see that Eq. (\ref{eqmo5}) corresponds to that for particle moving in external (1D) potential. The total energy of this ``particle'' can be written as                            \begin{equation}
\frac{{\dot z}^2}{2} + V(z)=0 \,.
\label{en1}
\end{equation}
with
\begin{equation}
V(z)= -2Az^2 + Bz^4 +2A-B \,.
\label{en2}
\end{equation}
where
\begin{equation}
A=2(JS)^2-f^2 \,, \quad B= 2(JS)^2 \,.
\label{en3}
\end{equation}
In Eqs. (\ref{en1}-\ref{en3}) we have again used the initial conditions $z(0)=\pm 1$ and ${\dot z}(0)=0$.

The behaviour of the ``potential eenergy'' $V(z)$ is presented schematically in Figure 5. For $f< JS\sqrt{2}$, the  potential is a double well, while for $f> JS\sqrt{2}$ the potential energy has a single minimum at $z=0$. Therefore point $f_c=JS\sqrt{2}$ corresponds to the phase transition discussed in the main text. Indeed, we note that the parameter $W=2\sqrt{\sum_j J_{ij}^2}$ corresponds in the LMG model, with the typical spin size (\ref{typical}), to the combination $ W=2J\sqrt{N} \approx 2\sqrt{2}J S$, and hence to the critical condition in LMG model given by
$$
f_c=W/2 \,,
$$
which reproduces the result of our estimates of the critical point in the main text.

\begin{figure}[t!]
\includegraphics[width=1.0\columnwidth,keepaspectratio]{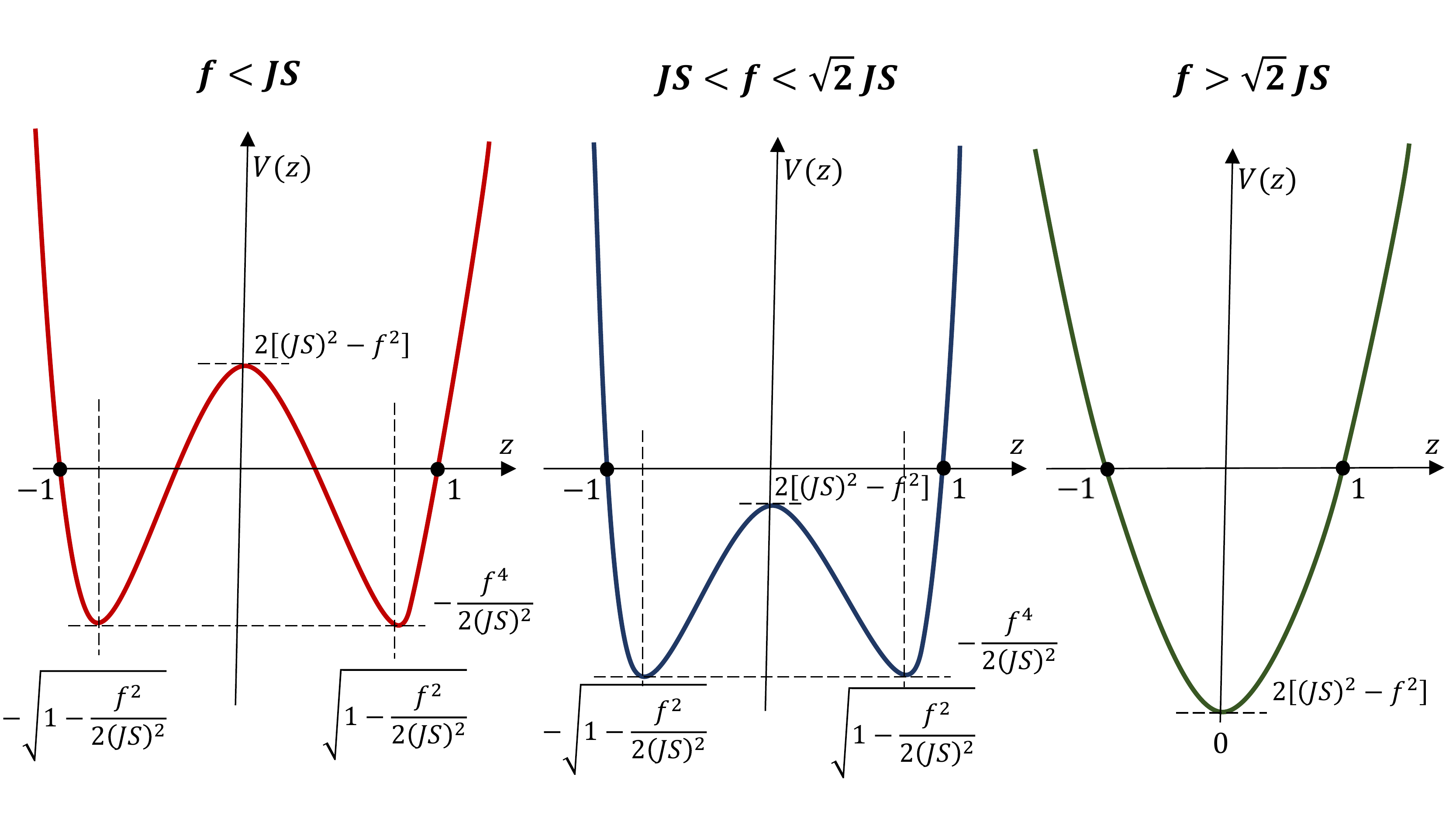}			
\caption{\label{fig5} Effective ``Ginzburg-Landau'' potentials for different phases.}
\end{figure}

Another important point corresponds to the value of $f$ being equal to $JS$, e.g. the red curve versus the blue curve in Fig.~5. Indeed, for $f$ smaller than $JS$, the ``particle'' starting at $z=\pm 1$ with zero initial velocity, cannot penetrate through the barrier and reach the other well. Such a transition is impossible (red curve in Fig.~5). For the blue curve such transition is possible. Therefore, for the dynamics in the potential given by the red curve, the quantity $z(0)z(t)$ does not change its sign, which leads to the divergence of the integral $\int dt e^{i\omega t} z(0)z(t)$ at $\omega=0$. This corresponds to results of the numerical simulation as well as to the theoretical results of the main text: for small enough $f$, the spectrum of the (z-component) spin fluctuations is divergent at $\omega=0$. For the dynamics along the blue curve $z(t)$ changes its sign and therefore the integral $\int dt z(0)z(t)$ is no longer divergent. This explains the crossover behaviour exhibited by the fluctuation spectrum in Fig.~1 and Eq.~(14) of the main text.

Finally, we must note that in the limit $N\rightarrow\infty$ the fluctuation spectrum is not going to exhibit a true phase transition (i.e., fully vanishing spectral density at and around $\omega=0$ for $f>W/2$), but a crossover. This is related to the fact that as $N\rightarrow\infty$ there are trajectories corresponding to the values of $S$ that violate the condition $f>\sqrt{2}JS$. For such trajectories the spectral weight at $\omega=0$ is finite (or even infinite). The probabilities for such trajectories for large enough $f$ are exponentially small, but finite from the formal standpoint (recall that $S$ is a random variable obeying Gaussian distribution). This is related to the infinite interaction range in the LMG model, which allows for its integrability (e.g. ``energy conservation'' for the effective coordinate $z$). We conjecture that the finite interaction range (between spins) would effectively keep $S$ finite, thus leading to a true (and finite) critical value for $f$, as observed in the numerical simulations; see, for example, the inset in Figure 1 of the main text. A detailed study of of this issue, however, lies out of the scope of this paper.


\section{Noise Spectrum: time-dependent mean-field analysis}.

In this section we present the missing details of the calculation of the spectrum of the spin auto-correlation function outlined in the main text.

Upon averaging over the rapidly varying fields $h_f(t)$, Eq. (10) of the main text can be written as
\begin{equation}\label{sig}
{\dot \sigma^z_s} = -4f^2 \int _0^t dt^\prime \sigma^z_s(t^\prime) F(t-t^\prime)\,,
\end{equation}
with
\begin{equation}\label{F}
F(\tau) =\cos(2h_s\tau)\exp\Big[-4\int \frac{d\omega}{2\pi}\,g_f(\omega)\frac{1-\cos(\omega\tau)}{\omega^2}\Big] .
\end{equation}

Note that the spectrum $g_f(\omega)$ can be associated with the Fourier transform of the averaged Larmor part of the spectrum for the correlation function $C^0(t)$ in Eq. (6) of the main text,
\begin{equation}\label{gf}
g_f^0(\omega) = \frac{W^2}{4}\int dh P(h) \frac{\pi f^2}{f^2+h^2}
\Big[\delta(\omega-\omega_h)+ \delta(\omega+\omega_h) \Big]  \,,
\end{equation}
where $\omega_h=2\sqrt{f^2+h^2}$. The superscript in $g_f^0(\omega)$ indicates that as far as the actual frequency dependence of the spectrum is concerned, this equation is a poor approximation: Indeed, due to the time dependence of $h$, the lines at $\pm\omega_h$ (i.e., the delta-functions) are broadened, which will clearly affect the shape of $g_h(\omega)$. Furthermore, $g_f^0(\omega)$ does not contain contributions from the cross-correlations between different spins. However, the variance of $h_f$,
$$
S_f=\int (d\omega/2\pi) g_f(\omega)\,,
$$
depends neither on the widths of the lines, nor on the spin cross-correlations, - the latter is equal to zero since different spins are uncorrelated at same times at infinite temperature. Therefore, while the shape of the spectrum given by Eq. (\ref{gf}) is a too crude an approximation, the estimate for its variance based on Eq. (\ref{gf}) is quite reasonable. This we obtain that
\begin{equation}\label{var}
S_f = \frac{W^2}{4}\int dh P(h) \frac{ f^2}{f^2+h^2}=\frac{W^2}{4}-h_0^2\,,
\end{equation}
where $h_0$ is given by Eq. (8) of the main text. In the following we will see that knowledge of the variance $S_f$ is sufficient for our purposes.

Solving Eq. (\ref{sig}) (supplemented by the initial condition $\sigma_s^z(0)=\pm 1$) by the method of Laplace transform, one readily obtains the power spectrum for the slow spins,
\begin{equation}\label{gs}
g_s(\omega, h_s) = \frac{2\Gamma(\omega, h_s)}{\big[\omega-\Omega(\omega, h_s)\big]^2 + \Gamma^2(\omega, h_s)},
\end{equation}
where $\Gamma$ and $\Omega$ are the real and imaginary parts of $2f^2\int_0^\infty d\tau e^{-i\omega\tau}F(\tau)$, respectively. This equation enters in Eq. (12) of the main text.

{\it Noise spectrum at small $\omega$ near the transition point:}

In the limit of vanishing $h_s$ and small $\omega$, functions $\Gamma(\omega,h_s)$ and $\Omega(\omega,h_s)$ in Eq. (\ref{gs}) can be approximated as $\Gamma(0,0)$ and $\omega\,\partial\Omega(0,0)/\partial\omega$, i.e..,
\begin{equation}\label{gam}
\Gamma\simeq 2f^2\int_0^\infty d\tau \exp\Big[-4\int \frac{d\omega}{2\pi}\,g_f(\omega)\frac{1-\cos(\omega\tau)}{\omega^2}\Big]\,,
\end{equation}
\begin{equation}\label{omega}
\Omega(\omega)\simeq 2\omega f^2\int_0^\infty d\tau \tau\exp\Big[-4\int \frac{d\omega}{2\pi}\,g_f(\omega)\frac{1-\cos(\omega\tau)}{\omega^2}\Big]\,.
\end{equation}
In evaluating these integrals we apply the saddle point method. While in the vicinity of the phase transition (i.e. for $f\sim W/2$) the accuracy of the method is poorly controlled (i.e., within $O(1)$, since the exponent does not contain a large parameter), such an approximation is consistent with the assumption that the distribution $P(h_s)$ is Gaussian (e.g. Eq. (7) of the main text). Indeed, the latter assumption is based on the central limit theorem, which also utilizes the saddle point method \cite{reif}.

The saddle point (in the integral over $\tau$ in Eq. (\ref{gam})) is at $\tau^\ast=0$. Expanding the function in the exponent in Eq. (\ref{gam}) around the saddle point, the resulting integral can be written as
\begin{equation}\label{gam1}
\Gamma\simeq 2f^2\int_0^\infty d\tau \exp(-2S_f\,\tau^2)\,,
\end{equation}
where the variance $S_f$ is given by Eq. (\ref{var}), $S_f\simeq W^2/4=f_c^2$. This yields
$$\Gamma\simeq \sqrt{2\pi} f_c\,.$$

In calculation of $\Omega(\omega,0)$ we apply the same approach. Then Eq. (\ref{omega}) can be approximated as
\begin{equation}\label{omega1}
\Omega(\omega)\simeq 2\omega f^2\int_0^\infty d\tau \tau\exp(-2S_f\,\tau^2) \,,
\end{equation}
which yields
$$\Omega(\omega)\simeq (f/f_c)^2\omega\,.$$

Substituting $\Gamma$ and $\Omega(\omega)$ into Eq. (\ref{gs}), we obtain the expression for the spectrum of slow spins,
\begin{equation}\label{gs1}
g_s(\omega) = \frac{2\sqrt{2\pi}f_c}{\delta^2\omega^2+2\pi f_c^2} \,,
\end{equation}
with $\delta$ given by Eq. (9) of the main text.

Substituting this expression into Eq. (12) of the main text and replacing the factor $h_s^2/(f^2+h_s^2)$ by $h_s^2/f_c^2$ (recall that $h_s\ll f$ in the vicinity of the transition point), we obtain
\begin{equation}\label{g1}
g(\omega) = \frac{2\sqrt{2\pi}f_c}{\delta^2\omega^2+2\pi f_c^2} \int dh_s P(h_s)\, \frac{h_s^2}{f_c^2} \,,
\end{equation}
with $P(h_s)$ given by Eq. (7) of the main text. Performing the remaining Gaussian integral, we obtain Eq. (13) of the main text.

{\it Noise spectrum at small $\omega$ for $f\ll f_c$:}

At sufficiently small $f$, the typical values of $h_s$ greatly exceed $f$ and therefore the relative fraction of slow spins, i e. the fraction in the integrand in Eq. (12) in the main text, can be replaced by $1$.
Furthermore, in the limit of small $f$ and at small enough $\omega$ (see below), we can neglect $\Omega(\omega,h_s)$ in Eq. (\ref{gs}) and set $\omega=0$ in the expression for $\Gamma(\omega,h_s)$. Then, Eqs. (12) of the main text can be simplified as
\begin{equation}\label{gs2}
g(\omega) = \int dh_s \,\frac{e^{-\frac{h_s^2}{2h_0^2}}}{\sqrt{2\pi h_0^2}}\, \frac{2\Gamma(h_s)}{\omega^2 + \Gamma^2(h_s)} \,,
\end{equation}
where $\Gamma(h_s)$ is equal to
\begin{equation}\label{gam0}
2f^2\int_0^\infty d\tau \cos{(2h_s\tau)}\exp\Big[-4\int \frac{d\omega}{2\pi}\,g_f(\omega)\frac{1-\cos(\omega\tau)}{\omega^2}\Big]\,.
\end{equation}
Writing cosine as a sum of two exponents of complex arguments, we can evaluate the integrals over $\tau$ by the saddle point method. Note that in the limit $f\ll W$ the exponent in the integrand function has a large parameter ($W/f$) and so the use of the saddle point method is well justified. The saddle points are given by the following equation,
\begin{equation}\label{sad0}
\pm i h_s = 2 \int \frac{d\omega}{2\pi}\, g_f(\omega)\,\frac{\sin(\omega\tau^\ast)}{\omega} \,.
\end{equation}
Assuming that the argument of sine in the above equation is small compared to $1$ (in magnitude), we find the saddle points at
\begin{equation}\label{sad}
\tau^\ast\simeq \pm i h_s/(2S_f)\,,
\end{equation}
with the variance $S_f$ given by Eq. (\ref{var}). In what follows we will see that the relevant values of $h_s$ are of the order of $\sqrt{Wf}$; see Eq. (\ref{gam1}) below. Also, from Eq. (\ref{var}) and Eq. (8) of the main text, one can readily verify that $S_f\simeq \sqrt{\pi/8}\, W f$ for $f\ll W$, and so $\tau^\ast\sim i/\sqrt{W f}$. Furthermore, from Eq. (\ref{gf}) as well as from the numerical noise spectra presented in Figure 1 of the main text, one can see that $g_f(\omega)$ decays at frequencies of the order of a few $f$. Then the quantity $\omega\tau^\ast$ in $\sin(\omega\tau^\ast)$ in Eq. (\ref{sad0}) is effectively bounded by $\sqrt{f/W}\ll 1$, and, therefore, the approximate solution for the saddle point in Eq. (\ref{sad}) is well justified. Then , after a straightforward calculation we obtain

\begin{equation}\label{gam1}
\Gamma(h_s)= f^2\sqrt{\frac{2\pi}{S_f}} \exp\Big(-\frac{h_s^2}{2S_f}\Big) \,.
\end{equation}
From Eq. (\ref{gam1} we see that the relevant values for $h_s$ are indeed bounded by $\sqrt{S_f}\sim \sqrt{Wf}$ for $f\ll W$ and, therefore, our calculation for $\Gamma(h_s)$ is internally consistent.

Substituting Eq. (\ref{gam1}) into Eq. (\ref{gs2}) one obtains the desired expression for the low frequency part of the fluctuation spectrum. It is, however, useful to change the integration variable in the resulting integral from $h_s$ to $\Gamma$. Then one obtains
\begin{equation}\label{g3}
g(\omega)\sim\int_0^{\Gamma(0)} d\Gamma \, \Big|\frac{dh_s(\Gamma)}{d\ln\Gamma}\Big| \,\frac{\Gamma^\alpha}{\omega^2 + \Gamma^2}\,,
\end{equation}
where we have dropped the proportionality constant to simplify notations. Here
\begin{equation}\label{al}
\alpha = \frac{W^2}{4h_0^2}-1\,.
\end{equation}
Note that for $f\ll W$, $\alpha\simeq \sqrt{2\pi}f/W$, and so $\alpha\ll 1$.

Since $h_s$ depends on $\Gamma$ as ``square root of logarithm'', within logarithmic accuracy one can assume that the factor $|dh_s/d ln\Gamma|$ in Eq. (\ref{g3}) is a constant. Therefore Eq. (\ref{g3}) reduces to Eq. (14) of the main text.


\begin{thebibliography}{100}

\bibitem{SNS} N. A. Sinitsyn, and Y. V. Pershin, ``The Theory of Spin Noise Spectroscopy: A Review", Rep. Prog. Phys. {\bf 79}, 106501 (2016).

\bibitem{MOF} C.-J. Yu, S. von Kugelgen, M. D. Krzyaniak, W. Ji, W. R. Dichtel, M. R. Wasielewski, D. E. Freedman,
``Spin and Phonon Design in Modular Arrays of Molecular Qubits", Chem. Mater. {\bf 32}, 23, 10200 (2020).

\bibitem{mol-nano} N. V. Prokof'ev and P. C. E. Stamp, ``Quantum Relaxation of Ensembles of Nanomagnets", Journal of Low Temperature Physics {\bf 113}, 1147 (1998).

\bibitem{purified} W. Wernsdorfer,``Quantum dynamics in molecular nanomagnets", Comptes Rendus Chimie
{\bf 11}, 1086 (2008).

\bibitem{ab} Similar spin Hamiltonian has been considered in connection with time crystals; see e.g., W. W. Ho, S. Choi, M. D. Lukin, D. A. Abanin, ``Critical Time Crystals in Dipolar Systems'', Phys. Rev. Lett. {\bf 119}, 010602 (2017).

\bibitem{SNS-exp} S. A. Crooker, D. G. Rickel, A. V. Balatsky, and D. L. Smith, ``Spectroscopy of spontaneous spin noise as a probe of spin dynamics and magnetic resonance", Nature {\bf 431},  49-52 (2004).

\bibitem{SNS-exp-rev} V. S. Zapasskii, ``Spin-noise spectroscopy: from proof of principle to applications", Advances in Optics and Photonics {\bf 5},  131 (2013).

\bibitem{QD1}K. A. Al-Hassanieh, V. V. Dobrovitski, E. Dagotto, and B. N. Harmon, ``Numerical Modeling of the Central Spin Problem Using the Spin-Coherent-State P-Representation",
Phys. Rev. Lett. {\bf 97}, 037204 (2006).

\bibitem{QD2} A. Fischer, I. Kleinjohann, N. A. Sinitsyn, and F. B. Anders, ``Cross-correlation spectra in interacting quantum dot systems", Phys. Rev. B {\bf 105}, 035303 (2022).

\bibitem{suppl} Supplemental material, with additional references \cite{LL1,LMG}, describes our numerics and defines different tested distributions of $J_{ij}$. It also provides a discussion of the phase transition in exactly solvable Lipkin-Meshkov-Glick model, and describes self-consistent calculations of the power spectrum in the entire range of frequencies.

\bibitem{LL1} N. A. Sinitsyn, V. V. Dobrovitski, S. Urazhdin, A. Saxena, ``Geometric Control Over the Motion of Magnetic Domain Walls", Phys. Rev. B {\bf 77}, 212405 (2008).

\bibitem{LMG} Y. Huang, T. Li, Z. Yin,``Symmetry-breaking dynamics of the finite-size Lipkin-Meshkov-Glick model near ground state", Phys. Rev. A {\bf 97}, 012115 (2018).

\end{thebibliography}

\begin{thebibliography}{100}

\bibitem{LMG} Yi Huang, Tongcang Li, and Zhang-qi Yin,``Symmetry-breaking dynamics of the finite-size Lipkin-Meshkov-Glick model near ground state", Phys. Rev. A {\bf 97}, 012115 (2018).

\bibitem{LL1} N. A. Sinitsyn, V. V. Dobrovitski, S. Urazhdin, A. Saxena, ``Geometric Control Over the Motion of Magnetic Domain Walls", Phys. Rev. B {\bf 77}, 212405 (2008).

\bibitem{reif} F. Reif, ``Fundamentals of statistical and thermal physics'', McGraw-Hill (1965).

\end{thebibliography}
\end{document}